\documentclass[conference]{IEEEtran}
\usepackage{graphicx}
\usepackage{balance}
\usepackage{amsmath}
\usepackage{amssymb}
\usepackage{multicol}

\usepackage{epsf}
\usepackage{acronym}

\usepackage{epstopdf}
\usepackage{epsf}
\usepackage{fancyhdr}
\usepackage{hyperref}

\usepackage[usenames]{color}

\usepackage{nopageno}%
\providecommand{\U}[1]{\protect\rule{.1in}{.1in}}

\begin{document}

\title{Performance Analysis of Geographic Routing Protocols in Ad Hoc Networks}

%

\title
{Performance Analysis of Geographic Routing Protocols in Ad Hoc Networks}
\author{\IEEEauthorblockN{ Don Torrieri,\IEEEauthorrefmark{1}
Salvatore Talarico,\IEEEauthorrefmark{2}
and Matthew C. Valenti\IEEEauthorrefmark{2} } \IEEEauthorblockA
{\IEEEauthorrefmark{1}%
U.S. Army Research Laboratory, Adelphi, MD, USA \\ \IEEEauthorrefmark
{2}West Virginia University, Morgantown, WV, USA} }%
%

\maketitle
%

\begin{abstract}%

Geographic routing protocols greatly reduce
the requirements of topology storage and provide flexibility
in the accommodation of the dynamic behavior of ad hoc
networks. This paper presents performance evaluations and
comparisons of two geographic routing protocols and the
popular AODV protocol. The trade-offs among the average
path reliabilities, average conditional delays, average conditional number of hops, and area spectral efficiencies and the
effects of various parameters are illustrated for finite ad hoc
networks with randomly placed mobiles. This paper uses a
dual method of closed-form analysis and simple simulation
that is applicable to most routing protocols and provides
a much more realistic performance evaluation than has
previously been possible. Some features included in the
new analysis are shadowing, exclusion and guard zones,
and distance-dependent fading.

\end{abstract}%

\section{Introduction}

Mobile ad hoc networks often use the ad-hoc on-demand distance-vector (AODV)
routing protocol \cite{perk}, which discovers and maintains multihop paths
between source mobiles and destination mobiles. However, these paths are
susceptible to disruption due to changes in the fading, terrain, and
interference, and hence the control overhead requirements are high. An
alternative class of routing protocols that do not maintain established routes
between mobiles are the geographic routing protocols. These protocols require
only a limited amount of topology storage by mobiles and provide flexibility
in the accommodation of the dynamic behavior of ad hoc networks \cite{cad},
\cite{ghaf}.

Among the many varieties of geographic routing protocols, four
representative ones are evaluated in this paper: greedy forwarding and known
nearest-neighbor routing, which use beacons, and contention-based
nearest-neighbor and maximum-progress routing, which are beaconless. The
tradeoffs among the average path reliabilities, average conditional delays,
average conditional number of hops, and area spectral efficiencies and the
effects of various parameters are illustrated for large ad hoc networks with
randomly placed mobiles.. A comparison is made with the popular AODV routing
protocol to gain perspective about the advantages and disadvantages of
geographic routing.

This paper uses a dual method of closed-form analysis and simple simulation to
provide a realistic performance evaluation of the five routing protocols. The
method performs spatial averaging over network realizations by exploiting the
deterministic geometry of \cite{tor} rather than the conventional stochastic
geometry \cite{haen}, thereby eliminating many unrealistic restrictions and
assumptions, as explained in \cite{tor4}. The method has great generality and
can be applied to the performance evaluation of most other routing protocols.

\section{Network Model}

The network comprises $M+2$ mobiles in an arbitrary two- or three-dimensional
region. The variable $X_{i}$ represents both the $i^{th}$ mobile and its
location, and $||X_{j}-X_{i}||$ is the distance from the $i^{th}$ mobile to
the $j^{th}$ mobile. Mobile $X_{0}$ serves as the reference transmitter or
message source, and mobile $X_{M+1}$ serves as the reference receiver or
message destination. The other $M$ mobiles $X_{1},...,X_{M}$ are potentially
relays or sources of interference. Each mobile uses a single omnidirectional antenna.

\emph{Exclusion zones} surrounding the mobiles, which ensure a minimum
physical separation between two mobiles, have radii set equal to
$r_{\mathsf{ex}}.$ The mobiles are uniformly distributed throughout the
network area outside the exclusion zones, according to a \textit{uniform
clustering} model \cite{tor1}.

The mobiles of the network transmit asynchronous quadriphase direct-sequence
signals. For such a network, interference is reduced after despreading by the
factor $h/G$, where $G$ is the \emph{processing gain} or \emph{spreading
factor}, and $h$ is the chip factor \cite{tor}, which reduces interference
due to its asynchronism. Let ${P}_{i}$ denote the
received power from $X_{i}$ at the reference distance $d_{0}$ before
despreading when fading and shadowing are absent. After the despreading, the
power of $X_{i}$'s signal at the mobile $X_{j}$ is
\begin{equation}
\rho_{i,j}=\tilde{P}_{i}g_{i,j}10^{\xi_{i,j}/10}f\left(  ||X_{j}%
-X_{i}||\right)  \label{1}%
\end{equation}
where $\tilde{P}_{i}={P}_{i}$ for the desired signal, $\tilde{P}_{i}=h{P}%
_{i}/G$ for an interferer, $g_{i,j}$ is the power gain due to fading,
$\xi_{i,j}$ is a shadowing factor, and $f(\cdot)$ is a path-loss function. The
path-loss function is expressed as the power law
\begin{equation}
f\left(  d\right)  =\left(  \frac{d}{d_{0}}\right)  ^{-\alpha}\hspace
{-0.45cm},\,\,\text{ \ }d\geq d_{0} \label{2}%
\end{equation}
where $\alpha\geq2$ is the path-loss exponent, $d_{0}$\ is sufficiently far
that the signals are in the far field, and $r_{\mathsf{ex}}\geq d_{0}.$

The \{$g_{i,j}\}$ are independent with unit-mean but are not necessarily
identically distributed; i.e., the channels from the different $\{X_{i}\}$ to
$X_{j}$ may undergo fading with different distributions. For analytical
tractability and close agreement with measured fading statistics, Nakagami
fading is assumed, and $g_{i,j}=a_{i,j}^{2}$, where $a_{i,j}$ is Nakagami with
parameter $m_{i,j}$. It is assumed that the \{$g_{i,j}\}$ remain fixed for the
duration of a time interval but vary independently from interval to interval
(block fading).

In the presence of shadowing with a lognormal distribution, the $\{\xi
_{i,j}\}$ are independent zero-mean Gaussian random variables with variance
$\sigma_{s}^{2}$. For ease of exposition, it is assumed that the shadowing
variance is the same for the entire network, but the results may be easily
generalized to allow for different shadowing variances over parts of the
network. In the absence of shadowing, $\xi_{i,j}=0$. While the fading may
change from one transmission to the next, the shadowing remains fixed for the
entire session.

The \emph{service probability} $\mu_{i}$ is defined as the probability that
mobile $X_{i}$ can serve as a relay along a path from a source to a
destination, and $1-\mu_{i}$ is the probability that $X_{i}$ is a potential
interferer.  A mobile may not be able to
serve as a relay in a path from $X_{0}$ to $X_{M+1}$
because it
is already receiving a transmission, is already serving as
a relay in another path, is transmitting, or is otherwise
unavailable

With \emph{interference probability} $p_{i}$, a potentially interfering $X_{i}$
transmits in the same time interval as the desired signal. The $\{p_{i}\}$ can
be used to model the servicing of other streams, controlled silence, or failed
link transmissions and the resulting retransmission attempts. Mobiles $X_{0}$
and $X_{M+1}$ do not cause interference. When the mobile $X_{j}$ serves as a
potential relay, we set $p_{j}=0.$

Let $\mathcal{N}$ denote the noise power, and the indicator $I_{i}$ denote a
Bernoulli random variable with probability $P[I_{i}=1]=p_{i}$. Since the
despreading does not significantly affect the desired-signal power, (\ref{1})
and (\ref{2}) imply that the instantaneous signal-to-interference-and-noise
ratio (SINR) at the mobile $X_{j}$ for a desired signal from mobile
$X_{k}$ is
\begin{equation}
\gamma_{k,j}=\frac{g_{k,j}\Omega_{k,j}}{\displaystyle\Gamma^{-1}+\sum_{i=1,i\neq
k}^{M}I_{i}g_{i,j}\Omega_{i,j}}%
\end{equation}
where
\begin{equation}
\Omega_{i,j}=%
\begin{cases}
10^{\xi_{k,j}/10}||X_{j}-X_{k}||^{-\alpha} & i=k\\
\displaystyle\frac{h{P}_{i}}{GP_{k}}10^{\xi_{i,j}/10}||X_{j}-X_{i}||^{-\alpha}
& i\neq k
\end{cases}
\end{equation}
is the normalized power of $X_{i}$ at $X_{j}$, and $\Gamma=d_{0}^{\alpha}%
P_{k}/\mathcal{N}$ is the SNR when $X_{k}$ is at unit distance from $X_{j}$
and fading and shadowing are absent.

The \emph{outage probability} quantifies the likelihood that the interference,
shadowing, fading, and noise will be too severe for useful communications.
Outage probability is defined with respect to an SINR threshold $\beta$, which
represents the minimum SINR required for reliable reception. In general, the
value of $\beta$ depends on the choice of coding and modulation. An
\emph{outage} occurs when the SINR falls below $\beta$. In \cite{tor},
closed-form expressions are provided for the outage probability conditioned on
the particular network geometry and shadowing factors. Let $\boldsymbol{\Omega
}_{j}=\{\Omega_{0,j},...,\Omega_{M+1,j}\}$ represent the set of normalized
powers at $X_{j}$. Conditioning on $\boldsymbol{\Omega}_{j}$, the \emph{outage
probability} of the link from $X_{k}$ to receiver $X_{j}$ is
\begin{equation}
\epsilon_{k,j}=P\left[  \gamma_{k,j}\leq\beta\mid\boldsymbol{\Omega}_{j}\right]
.
\end{equation}
The conditioning enables the calculation of the outage probability for any
specific network geometry, which cannot be done using tools based on
stochastic geometry. The closed-form equations for $\epsilon_{k,j}$ are used
in the subsequent performance evaluations of the routing protocols.

\section{Routing Models}

\subsection{Routing Protocols}

The three routing protocols that are considered are
reactive or on-demand protocols that only seek routes
when needed and do not require mobiles to store details about large portions of the network. The AODV protocol relies on flooding to seek the
\emph{fewest-hops path} during its
\emph{path-discovery phase}. The flooding diffuses
request packets simultaneously over multiple routes for
the purpose of discovering a successful route to the destination despite link failures along some potential paths.
When the first request packet reaches the destination,
backtracking by an acknowledgement packet establishes
the route the request packet followed as the single
static fewest-hops path for subsequent message packets
during a \emph{message-delivery phase}. Subsequent receptions
of request packets by the destination are ignored. There
is a high overhead cost in establishing the fewest-hops
path during the path-discovery phase, and the fewest-
hops path must be used for message delivery before
changes in the channel conditions cause an outage of
one or more of its links.

Geographic protocols limit information-sharing costs
by minimizing the reliance of mobiles on topology
information \cite{cad},
\cite{ghaf}. Since geographic routing protocols
make routing decisions on a hop-by-hop basis, they do
not require a flooding process for path discovery. Two
geographic routing protocols are examined: the
\emph{greedy
forwarding protocol}
and the
\emph{maximum progress protocol}.
Both geographic routing protocols assume that each
mobile knows its physical location and the direction
towards the destination.

The greedy forwarding protocol relies on
\emph{beacons}, which are mobiles that periodically broadcast information about their locations. A source forwards a packet to a
\emph{relay}
that is selected from a set of neighboring
beacons that are modeled as the set of active mobiles that
lie within a
\emph{transmission range}
of radius
$r_t$. The next
link in the path from source
$X_{0}$ to destination
$X_{M+1}$ is the link to the relay within the transmission range
that shortens the remaining distance to
$X_{M+1}$ the most.
There is no path-discovery phase because the relays
have the geographic information necessary to route the
messages to the destination.

The maximum progress protocol is a contention-based
protocol that does not rely on beacons but comprises
alternating
\emph{path-discovery phases}
and
\emph{message-delivery
phases}. During a path-discovery phase, a single link to a
single relay is discovered. During the following message-delivery phase, a packet is sent to that relay, and then
the alternating phases resume until the destination is
reached. In a path-discovery phase, the next relay in a
path to the destination is dynamically selected at each
hop of each packet and depends on the local configuration of available relays. A source or relay broadcasts
\emph{Request-to-Send}
(RTS) messages to neighboring mobiles
that potentially might serve as the next relay along the
path to the destination. The RTS message includes the
location of the transmitting source or previous relay.
Upon receiving the RTS, a neighboring mobile initiates
a timer that has an expiration time proportional to the
remaining distance to the destination. When the timer
reaches its expiration time, the mobile sends a
\emph{Clear-to-Send}
(CTS) message as an acknowledgement packet to
the source or previous relay. The earliest arriving CTS
message causes the source or previous relay to launch
the message-delivery phase by sending message packets
to the mobile that sent that CTS message, and all other
candidate mobiles receiving that CTS message cease
operation of their timers.

\subsection{Implementation of Path Selection}

For the analysis and simulation, we draw a random
realization of the network (topology) using the uniform clustering distribution of mobiles. The source and
destination mobiles are placed, and then, one by one,
the location of each remaining
$X_i$ is drawn according to a uniform distribution within the network region. However, if an
$X_i$ falls within the exclusion zone of
a previously placed mobile, then it has a new random
location assigned to it as many times as necessary until
it falls outside all exclusion zones. Using the service
probabilities, the set of potential relays is randomly
selected for each simulation trial.

The routing protocols use a \emph{distance criterion} to exclude a link from
mobile $X_{i}$ to mobile $X_{j}$ as a link in one of the possible paths from
$X_{0}$ to $X_{M+1}$ if $||X_{j}-X_{M+1}||>||X_{i}-X_{M+1}||$. These
exclusions ensure that each possible path has links that always
reduce the remaining distance to the destination. All links connected to mobiles that cannot serve as relays are
excluded as links in possible paths from $X_{0}$ to $X_{M+1}.$ Links that have
not been excluded are called \emph{eligible} links.

The eligible links are used to determine the greedy-forwarding path from
$X_0$ to $X_{M+1}$
during its message-delivery phase. There is no path-discovery phase. If no
path from $X_0$ to $X_{M+1}$can be found or if the message
delivery fails, a \emph{routing failure} is recorded.

A \emph{candidate link} is an eligible link that does not experience an outage during the path-discovery phase. To identify the candidate links within each topology, we apply our analysis to determine the outage probability for each eligible link. A Monte Carlo simulation decides
whether an eligible link is in an outage by sampling a
Bernoulli random variable with the corresponding outage
probability. A links that is not in an outage is called a
\emph{candidate link}.

For AODV, the \emph{candidate paths} from $X_{0}$ to $X_{M+1}$ are paths that
can be formed by using candidate links. The candidate path with the fewest hops from $X_{0}$ to $X_{M+1}$
is selected as the
\emph{fewest-hops path}. This path is
determined by using the
\emph{Djikstra algorithm}
\cite{bru} with the
unit cost of each candidate link. If two or more candidate
paths have the fewest hops, the fewest-hops path is
randomly selected from among them. If there is no set
of candidate links that allow a path from
$X_{0}$ to $X_{M+1},$
then a routing failure occurs. If a fewest-hops path exists,
then a Monte Carlo simulation is used to determine
whether the acknowledgement packet traversing the path
in the reverse direction is successful. If it is not or if the
message delivery over the fewest-hops path fails, then a
routing failure occurs

A \emph{two-way candidate link}
is an eligible link that does
not experience an outage in either the forward or the
reverse direction during the path-discovery phase. A
Monte Carlo simulation is used to determine the two-way
candidate links. For the maximum progress protocol, the
two-way candidate link starting with source
$X_0$ with a terminating relay that minimizes the remaining distance
to destination $X_{M+1}$
is selected as the first link in the
maximum-progress path. The link among the two-way
candidate links that minimizes the remaining distance
and is connected to the relay at the end of the previously
selected link is added successively until the destination
$X_{M+1}$ is reached and hence the maximum-progress path
has been determined. After each relay is selected, a
message packet is sent in the forward direction to the
selected relay. If no maximum-progress path from
$X_0$ to $X_{M+1}$ can be found or if a message delivery fails, a
routing failure is recorded.

The CTS message transmitted by the maximum
progress protocol during its path-discovery phase establishes guard zones \cite{tor1}. Potentially interfering mobiles
within the guard zones are silenced during the message-delivery phase of the maximum progress protocol. It is assumed that the guard zones have sufficiently small radii
$r_g$ that the CTS message is correctly decoded. Any
potentially interfering mobile $X_i$
that lies in one of the guard zones surrounding the two mobiles at the ends of each link of a selected path is silenced by setting its
$p_{i}=0$ during message delivery.

\subsection{Performance Metrics}

Let $B$ denote the maximum number of transmission
attempts over a link of the path. During the path-discovery phases, $B=1$. During the message-delivery phases, $B\geq1$ because message retransmissions over an established link are feasible. For each eligible or
candidate link $l=(i,j)$, a Bernoulli random variable
with failure probability $\epsilon_{l}$
is repeatedly drawn until there
are either $B$ failures or success after
$N_l$ transmission attempts, where $N_l  \leq B$. The
\emph{delay of link}
$l$ of the
selected path is $N_l T +(N_l-1)T_e$, where $T$
is the \emph{delay of a transmission over a link}, and
$T_e$ is the \emph{excess delay} caused by a retransmission.

Each network topology $t$ is used in $K_t$ simulation trials. The
path delay $T_{s,t}$ of a path from
$X_0$ to $X_{M+1}$ for network topology $t$
and simulation trial $s$ is the sum
of the link delays in the path during the message-delivery
phase:
\begin{equation}
T_{s,t}=\sum\limits_{l\in\mathcal{L}_{s,t}}[N_{l}T+(N_{l}-1)T_{e}]
\end{equation}
where $\mathcal{L}_{s,t}$ is the set of links constituting the path.
If there are $B$ transmission failures for any link of the
selected path, then a routing failure occurs.

If there are $F_{t}$ routing failures for topology $t$ and $K_{t}$ simulation
trials, then the \emph{probability of end-to-end success} or \emph{path}
\emph{reliability} within topology $t$ is
\begin{equation}
R_{t}=1-\frac{F_{t}}{K_{t}}. \label{13}%
\end{equation}

Let $\mathcal{T}_{t}$ denote the set of
$K_{t}-F_{t}$ trials with no routing failures. If the selected path for trial $s$ has $h_{s,t}$ links or hops, then among the set
$\mathcal{T}_{t}$, the average conditional
\emph{number of hops}
from $X_{0}$ to $X_{M+1}$ is
\begin{equation}
H_{t}=\frac{1}{K_{t}-F_{t}}\sum\limits_{s\in\mathcal{T}_{t}}h_{s,t}.
\end{equation}

Let ${T}_{d}$ denote the link delay of packets during the path-discovery phase. The average conditional
\emph{delay} from $X_{0}$ to $X_{M+1}$
during the combined path-discovery and
message-delivery phases is
\begin{equation}
D_{t}=\frac{1}{K_{t}-F_{t}}\sum\limits_{s\in\mathcal{T}_{t}}\left(T_{s,t} + 2 c h_{s,t} T_{d}\right).
\end{equation}
where $c=0$ for the greedy forwarding protocol, and $c=1$ for the maximum progress and AODV protocols

Let $A$ denote the network area and $\lambda=(M+1)/A$ denote the density of
the possible transmitters in the network. We define the \emph{normalized}
\emph{area spectral efficiency} for the $K_{t}$ trials of topology $t$ as
\begin{equation}
\mathcal{A}_{t}=\frac{\lambda}{K_{t}}\sum\limits_{s=1}^{K_{t}}\frac{1}%
{T_{s,t}+ 2 c h_{s,t} T_{d}}%
\end{equation}
where the normalization is with respect to the bit rate or bits per channel use. The normalized area spectral
efficiency is a measure of the end-to-end throughput in the network. After
computing $R_{t},$ $D_{t},$ $H_{t},$ and $\mathcal{A}_{t}$ for $\Upsilon$
network topologies, we can average over the topologies to compute the \emph{topological averages: }$\overline{R,}$\ $\overline{D},\overline{\text{
}H},$ and $\overline{\mathcal{A}}.$

\section{Numerical Results}

A host of network topologies and parameter values can be evaluated by the
method described. Here, we consider a representative example that illustrates
the tradeoffs among the routing protocols. We consider a network occupying a
circular region with normalized radius $r_{net}=1.$ The source mobile is
placed at the origin, and the destination mobile is placed a distance
$||X_{M+1}-X_{0}||$ from it. Times are normalized by setting $T=1$. Each
transmitted power ${P}_{i}$ is equal. There are no retransmissions during the
path-discovery phases, whereas $B=4$ during the message-delivery phases$.$
A \emph{distance-dependent fading} model is assumed, where a signal
originating at mobile $X_{i}$ arrives at mobile $X_{j}$ with a Nakagami fading
parameter $m_{i,j}$ that depends on the distance between the mobiles. We set
\begin{equation}
m_{i,j}=%
\begin{cases}
3 & \mbox{ if }\;||X_{j}-X_{i}||\leq r_{\mathsf{f}}/2\\
2 & \mbox{ if }\;r_{\mathsf{f}}/2<||X_{j}-X_{i}||\leq r_{\mathsf{f}}\\
1 & \mbox{ if }\;||X_{j}-X_{i}||>r_{\mathsf{f}}%
\end{cases}
\end{equation}
where $r_{\mathsf{f}}$ is\ the \emph{line-of-sight radius}. The
distance-dependent-fading model characterizes the typical situation in which
nearby mobiles most likely are in each other's line-of-sight, while mobiles
farther away from each other are not. Other fixed parameter values are
$r_{ex}=0.05,$ $r_{\mathsf{f}}=0.2,$ $T_{e}=1.2,$ $T_{d}=1,$ $M=200$, $\beta=0$ dB,
$r_{\mathsf{g}}=0.15,$ $K_{t}=10^{4},$ $\Gamma=0$ dB, $\alpha=3.5,$
and $\Upsilon=2000.$ The service and interference probabilities
are assumed to have the same values for all mobiles
so that $\mu_{i}=\mu$ and
$p_{i}=p$. Unless otherwise stated, $G/h=96$, $\alpha=3.5$, $\mu=0.4,$ and $p=0.3.$ When
shadowing is present, it has a lognormal distribution with $\sigma_{s}=8$ dB.
However, the transmitted packets encounter the same shadowing in both directions over the same link during both routing phases.

\begin{figure}[tb]%
\centering
\includegraphics[width=9cm]{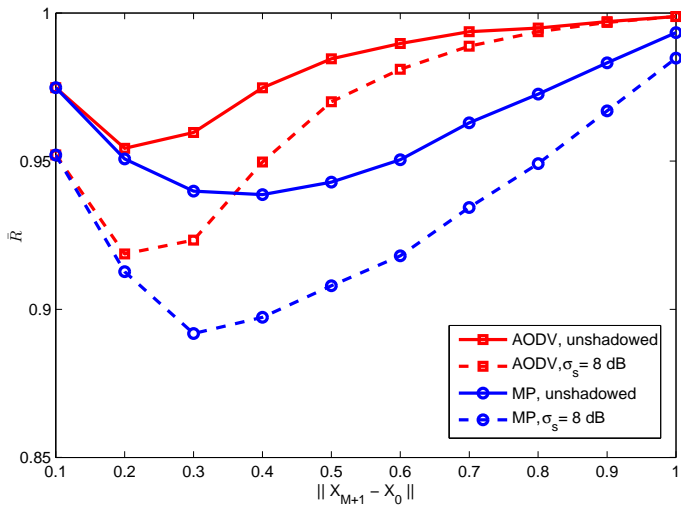}%
\caption{Average path reliability for request packets of AODV and MP protocols as a function of the distance between source and destination.} \label{Fig.1}
\end{figure}

\begin{figure}[tb]%
\centering
\includegraphics[width=9cm]{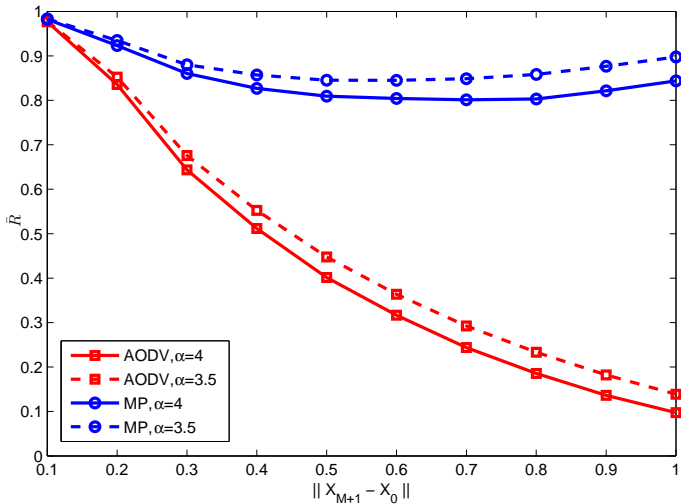}%
\caption{Average path reliability for acknowledgements of AODV and MP protocols as a function of the distance between source and destination.} \label{Fig.2}
\end{figure}

Fig. \ref{Fig.1} and Fig. \ref{Fig.2} display the average path reliabilities of the request packets
and acknowledgement packets, respectively, for the complete selected paths
during the path-discovery phases of the AODV and maximum
progress (MP) protocols. Figure 1 depicts the reliabilities both with and without
shadowing as a function of the source-destination distance $||X_{M+1}-X_{0}||.$ Shadowing is
assumed in Fig. \ref{Fig.2} and all subsequent figures. Fig. \ref{Fig.1} shows an initial decrease and then an increase in average path reliability as the source-destination distance
increases. This variation occurs because at short distances, there are very few relays that provide forward
progress, and often the only eligible or candidate link
is the direct link from source to destination. As the
distance increases, there are more eligible and candidate
links, and hence the network benefits from the diversity.
Furthermore, as the destination approaches the edge
of the network, the path benefits from a decrease in
interference at the relays that are close to the destination.
Fig. \ref{Fig.1} shows that during the request stage, the AODV
protocol provides the better path reliability because it
constructs several partial paths before the complete path
is determined.

Since the relays are already determined in Fig. \ref{Fig.2}, the
maximum progress protocol shows only a mild improvement with increasing source-destination distance, and
this can be attributed almost entirely to the edge effect. It
is observed in Fig. \ref{Fig.2} that the AODV protocol has a relatively poor path reliability during the acknowledgement
stage, which is due to the fact that a specified complete
path must be traversed in the reverse direction, where the
interference and fading may be much more severe. The
maximum progress protocol does not encounter the same
problem because the links in its paths are selected one-by-one with the elimination of links that do not provide
acknowledgements. Although both the shadowing and
the path-loss exponent $\alpha$ affect both the packets and
the interference signals, the two figures indicate that the
overall impact of more severe propagation conditions is
detrimental for all distances.

\begin{figure}[tb]%
\centering
\includegraphics[width=9cm]{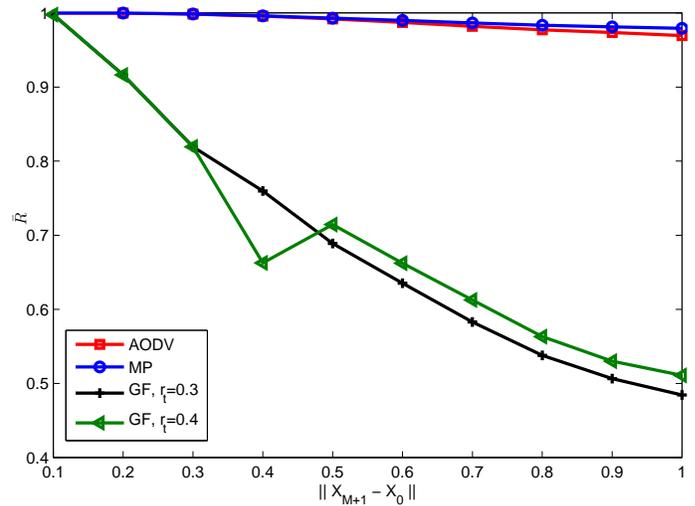}%
\caption{Average path reliability for message-delivery phase of each routing
protocol as a function of the distance between source and destination.} \label{Fig.3}
\end{figure}

Fig. \ref{Fig.3} displays the average path reliabilities for
the message-delivery phases of the three protocols, assuming that the path-discovery phase, if used, has been
successful. The figure illustrates the penalties incurred
by the greedy forwarding (GF) protocol because of
the absence of a path-discovery phase that eliminates
links with excessive shadowing, interference, or fading
and creates guard zones for the message-delivery phase.

The figure illustrates the role of the transmission range
$r_{t}$ in determining average path reliability for greedy
forwarding protocols. As $r_t$ increases, the links in the
complete path are longer and less reliable. However, this
disadvantage is counterbalanced by the increased number
of potential relays and the reduction in the average
number of links in a complete path.

\begin{figure}[tb]%
\centering
\includegraphics[width=9cm]{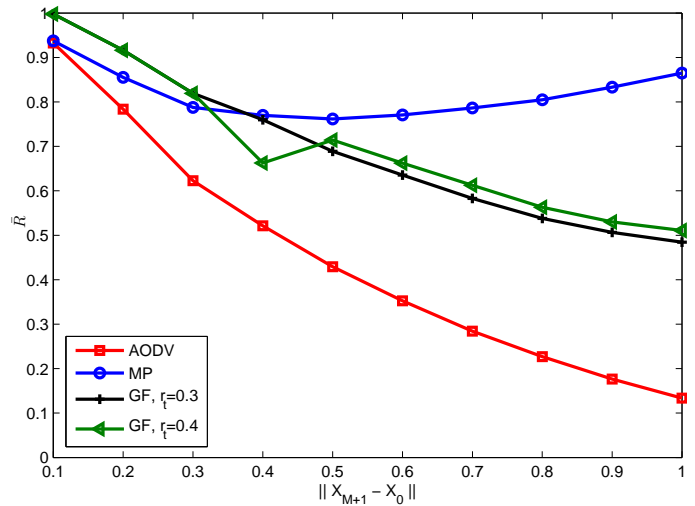}%
\caption{Average path reliability for both phases of each routing protocol as a function of the distance between source and destination.} \label{Fig.4}
\end{figure}

Fig. \ref{Fig.4} shows the overall average path reliabilities
for the combined path-discovery and message-delivery
phases of all three routing protocols. The AODV protocol
is the least reliable. The maximum progress protocol is
much more reliable than the greedy forwarding protocol
if $||X_{M+1}-X_{0}||$
is large, but is not as reliable if $||X_{M+1}-X_{0}||<0.35$
because of the relatively low reliability of its request packets.

\begin{figure}[tb]%
\centering
\includegraphics[width=9cm]{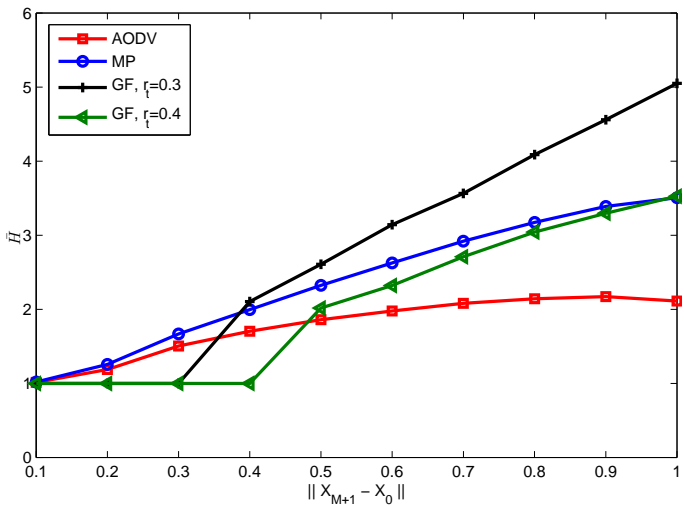}%
\caption{Average conditional delay of each routing protocol as a function of
the distance between source and destination. } \label{Fig.5}
\end{figure}

\begin{figure}[tb]%
\centering
\includegraphics[width=9cm]{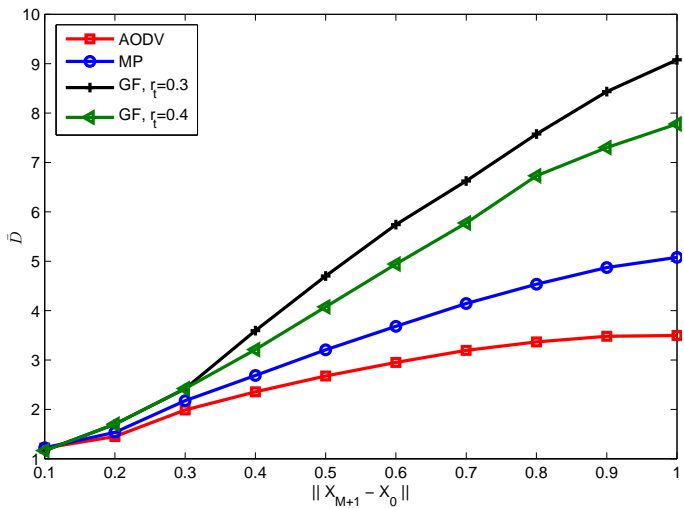}%
\caption{Average conditional number of hops of each routing protocol as a
function of the distance between source and destination.} \label{Fig.6}
\end{figure}

\begin{figure}[ptb]%
\centering
\includegraphics[width=9cm]{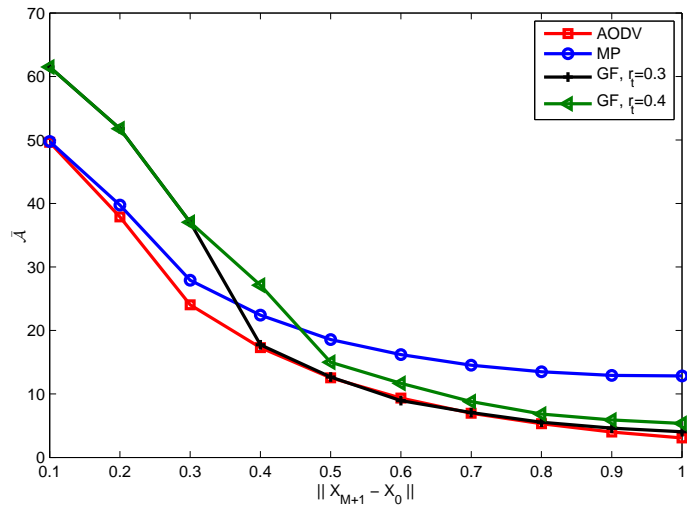}%
\caption{Area spectral efficiency of each routing protocol as a function of
the distance between source and destination.} \label{Fig.7}
\end{figure}

The average conditional delay $\overline{D},$ the average conditional number
of hops $\overline{H,}$ and the normalized area spectral efficiency
$\overline{\mathcal{A}}$ for each routing protocol as a function of
$||X_{0}-X_{M+1}||$ are displayed in Fig. \ref{Fig.5}, Fig. \ref{Fig.6}, and Fig. \ref{Fig.7}, respectively. The greedy forwarding protocol
has the highest $\overline{\mathcal{A}}$ if $||X_{M+1}-X_{0}||$
is small, whereas the maximum progress protocol has the highest $\overline{\mathcal{A}}$ if $||X_{M+1}-X_{0}||$ is large. The reason is the rapid loss of reliability and increase in the average conditional delay
of the greedy forwarding protocol when
$||X_{M+1}-X_{0}||$ is large.

\section{Conclusions}

This paper presents performance evaluations and comparisons of two geographic routing protocols and the
popular AODV protocol. The trade-offs among the average path reliabilities, average conditional delays, average
conditional number of hops, and area spectral efficiencies
and the effects of various parameters have been shown
for a typical ad hoc network. Since acknowledgements
are often lost due to the nonreciprocal interference and
fading on the reverse paths, the AODV protocol has a
relatively low path reliability, and its implementation
is costly because it requires a flooding process. In
terms of the examined performance measures, the greedy
forwarding protocol is advantageous when the separation
between the source and destination is small and the
spreading factor is large, provided that the transmission
range and the relay density are adequate. The maximum
progress protocol is more resilient when the relay density
is low and is advantageous when the separation between
the source and destination is large.

The general methodology of this paper can be used
to provide a significantly improved analysis of multihop
routing protocols in ad hoc networks. Many unrealistic
and improbable assumptions and restrictions of existing
analyses can be discarded.

\end{document}